\documentclass[runningheads]{llncs}
\usepackage[T1]{fontenc}
\usepackage{enumitem}
\usepackage{amsmath}
\usepackage{graphicx}
\graphicspath{{figures}}
\usepackage{booktabs}
\usepackage{adjustbox}
\usepackage{multirow}
\usepackage{color}

\begin{document}
\title{OpenCUBE: Building an Open Source Cloud Blueprint with EPI Systems\thanks{OpenCUBE is funded by European Commission Horizon Project 101092984.}}
%
%
\author{
Ivy Peng\thanks{Corresponding author ipeng@acm.org.}\inst{1} \and
Martin Schulz \inst{2} \and
Utz-Uwe Haus \inst{3}\and
Craig Prunty\inst{4} \and
Pedro Marcuello\inst{5} \and
Emanuele Danovaro\inst{6} \and
Gabin Schieffer\inst{1} \and
Jacob Wahlgren\inst{1} \and
Daniel Medeiros\inst{1} \and
Philipp Friese\inst{2} \and
Stefano Markidis \inst{1}
}

\authorrunning{Peng et al.}
\institute{
KTH Royal Institute of Technology, Sweden \and 
Technical University of Munich, Germany \and
Hewlett Packard Enterprise, Switzerland \and 
SiPearl, France \and 
Semidynamics, Spain \and
ECMWF, Italy
}

\maketitle              
\begin{abstract}
OpenCUBE aims to develop an open-source full software stack for Cloud computing blueprint deployed on EPI hardware, adaptable to emerging workloads across the computing continuum. OpenCUBE prioritizes energy awareness and utilizes open APIs, Open Source components, advanced SiPearl Rhea processors, and RISC-V accelerator. The project leverages representative workloads, such as cloud-native workloads and workflows of weather forecast data management, molecular docking, and space weather, for evaluation and validation.
\end{abstract}

\section{Introduction}
OpenCUBE is a project funded by the European Commission and initiated in January 2023. Its primary aim is to create, implement, and validate a full software stack for enabling a European Cloud computing blueprint deployed on European hardware infrastructure and cater to industrial and consumer cloud workloads. Additionally, the project aims to prioritize power and energy efficiency by incorporating power awareness at all levels. OpenCUBE will support the diverse requirements of the entire computing continuum, spanning from edge to cloud and high-performance computing (HPC).

OpenCUBE involves designing and installing a prototype hardware infrastructure composed of SiPearl processors and Semidynamics RISC-V accelerators, which are outcomes of the European Processor Initiative (EPI). Heterogeneous compute nodes will be interconnected with a high-performance Ethernet network to support the exploration of emerging memory and storage disaggregation. OpenCUBE will create a unified software stack encompassing various best practices and open-source tools at the operating system, middleware, and system management levels. The OpenCUBE software stack for cloud services will be open-source and leverage industry-standard Open APIs and Open Source components. 
 
In line with the European Green Deal initiative, OpenCUBE is designed to enable energy awareness as a fundamental feature across the entire stack. Through software-hardware co-design, the OpenCUBE software stack will provide API access to various site levels, from core, socket, node, and even to the electricity grid. The project will utilize representative workloads, such as weather forecast data management, molecular docking, and space weather workflows, to inform the design and deployment of the OpenCUBE system.

\section{Approach}
\begin{figure}
    \centering
    \includegraphics[width=\linewidth]{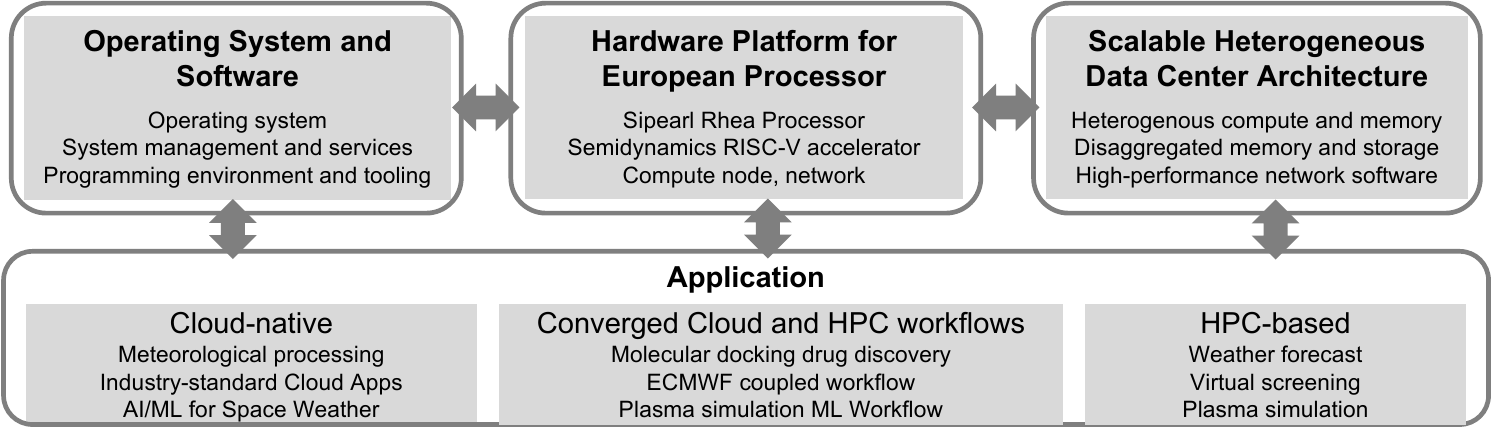}
    \caption{An overview of the OpenCUBE approach with a focus on close interface between software and hardware.}
    \label{fig:overview}
    \vspace{-10pt}
\end{figure}

OpenCUBE takes an approach focusing on close hardware and software interfaces and is organized into four thrusts -- hardware platform, middleware, heterogeneous data center (DC), and applications (Figure~\ref{fig:overview}).

\noindent$\bullet$~~\textbf{The Hardware Platform} is designed for hosting EPI systems, e.g., SiPearl Rhea and Semidynaimcs RISC-V processors. The design is adapted for catering to both cloud-native and HPC workloads' requirements. Also, the hardware platform provides inputs to the design of the operating system and middleware for a heterogeneous data center. As in the early phase, a prototype platform is built to exploit ARM-based processors available in commercial cloud systems, high-performance Ethernet-based interconnects, and the EPAC RISC-V accelerator.

\noindent$\bullet$~~\textbf{Operating Systems and Software} will leverage and extend open-source components to support newly released hardware features for monitoring and resource management on the hardware platform. Extensions to system management services and OS will be designed to utilize the high-performance Ethernet network and enable power awareness at various system levels to achieve energy efficiency and performance. Open-source profiling tools will be adapted to the new CPU architecture to improve performance tuning and debugging.

\noindent$\bullet$~~\textbf{A Scalable Heterogeneous Data Center} architecture will be deployed through open-source middleware specifically adapted to the hardware platform and targeted workloads. In particular, the OpenCUBE middleware will focus on improving the utilization of heterogeneous compute and memory resources on node. Also, extensions are to be developed to enable memory disaggregation over the fast network for either memory expansion or data staging in workflows. High-performance network software is designed to enable the efficient execution of applications in cloud-based containerized environments. For instance, MPI is the dominant communication API for traditional HPC applications. However, its relatively static model, which was developed for conventional HPC setups and schedulers, needs to be revised for and adapted to cloud-based application deployment, for example, using the recently introduced MPI Sessions concept~\cite{holmes2016mpi}. 

\noindent$\bullet$~~\textbf{Driver Applications \& Workflows} will interact with hardware, software, and DC middleware thrusts to input workload requirements. The OpenCUBE stack will be validated and evaluated through these applications, including cloud-native workloads and workflows targeting the computing continuum from cloud, HPC, and edge. The applications will continuously provide feedback throughout the development of the stack. For instance, ECMWF~\cite{bauer2015quiet}’s IFS will drive customization of general data storage middleware. A workflow with integrated ML-based analytics and iPIC3D~\cite{markidis2010multi}-based simulation will guide the software stack for converged HPC and cloud. A virtual screening workflow based on AutoDock~\cite{santos2021accelerating}, a widely used molecular docking software for drug discovery, will be used for validating cloud-based workflows on heterogeneous resources.

\section{Preliminary Results and Roadmap}
In the first phase, we are deploying a prototype hardware platform of one rack of four HPE ProLiant RL300 servers with 256 GB DDR4 memory and a minimum of 1 TB NVMe SSD for provisioning fabric-attached memory. An FPGA emulator is employed to integrate EPAC RISC-V processors for acceleration. The platform is equipped with Slingshot interconnects, including one Cassini Switch GB Ethernet and the Cassini network interface card. 

To enable cloud-native workflows, we investigate a popular workflow management software, Apache Airflow. As a case study, we used a virtual screening software in drug discovery, AutoDock, to enable a workflow of automatic elastic molecular docking on the cloud~\cite{medeiros2023airflow}. Our preliminary results confirm the feasibility of deployment into the containerized environment. We investigated the state-of-the-art disaggregated memory technologies such as Compute Express Link (CXL) for the scalable heterogeneous data center architecture. We developed a memory-centric profiling tool and a software emulation framework to explore design space~\cite{wahlgren2023mem} quantitatively. 

To enable converged computing between Cloud and HPC, we developed a framework atop the open-source container orchestrator Kubernetes that enables the reuse of already provisioned infrastructure. This capability enables automatic horizontal scaling for tightly coupled MPI-based applications, which is cumbersome to realize on traditional HPC systems~\cite{medeiros2023elastic}. In analyzing application requirements for input to OpenCUBE stack design, we also identified a scalability bottleneck in reduction operation due to a large number of synchronization points. We proposed a matrix-based multi-dimensional reduction algorithm for accelerating the local search of the scoring function and explored a tensor-based implementation for optimizing the molecular docking process~\cite{schieffer2023tcu}. The results show an over 25\% improvement in average docking time for a real-world docking scenario. 

On the roadmap towards an open-source cloud blueprint, the OpenCUBE project employs an approach focusing on close interaction between software and hardware development to create an open-source cloud blueprint on EPI systems. The prototype implementation in OpenCUBE will be validated and evaluated with industrial and consumer cloud applications. Development in enabling adaptive MPI communication setup, e.g., session, will also provide feedback to standardization bodies~\cite{mpi4}. OpenCUBE's roadmap aligns the major upgrade of the prototype hardware infrastructure with chips produced from EPI. Meanwhile, insights and findings learned during the design and development of the OpenCUBE stack based on the Sipearl Rhea and Semidynamics RISC-V processors are feedback to the Open Source, EPI, and Computing Continuum communities.\\

\noindent\textbf{Acknowledgment:} OpenCUBE is funded by EU Horizon Project 101092984.

\bibliographystyle{splncs04}
\bibliography{main}

\end{document}